# Quantitative Nanoscale Mapping of Three-Phase Thermal Conductivities in Filled Skutterudites via Scanning Thermal Microscopy

Ehsan Nasr Esfahani[*,†,a], Feiyue Ma[*,a], Shanyu Wang[‡], Yun Ou[†], Jihui Yang[‡], and Jiangyu Li[*,†,§]

## ABSTRACT

In the last two decades, a nanostructuring paradigm has been successfully applied in a wide range of thermoelectric materials, resulting in significant reduction in thermal conductivity and superior thermoelectric performance. These advances, however, have been accomplished without directly investigating the local thermoelectric properties, even though local electric current can be mapped with high spatial resolution. In fact, there still lacks an effective method that links the macroscopic thermoelectric performance to the local microstructures and properties. Here, we show that local thermal conductivity can be mapped quantitatively with good accuracy, nanometer resolution, and *one-to-one* correspondence to the microstructure using a three-phase skutterudite as a model system. Scanning thermal microscopy combined with finite element simulations demonstrate close correlation between sample conductivity and probe resistance, enabling us to distinguish thermal conductivities spanning orders of magnitude, yet resolving thermal variation across a phase interface with small contrast. The technique thus provides a powerful tool to correlate local thermal conductivities, microstructures, and macroscopic properties for nanostructured materials in general, and nanostructured thermoelectrics in particular.

[*] Department of Mechanical Engineering, University of Washington, Seattle, WA 98195, USA.
[†] Shenzhen Key Laboratory of Nanobiomechanics, Shenzhen Institutes of Advanced Technology, Chinese Academy of Sciences, Shenzhen 518055, Guangdong, China
[‡] Department of Materials Science and Engineering, University of Washington, Seattle, WA 98195, USA.
[§] To whom the correspondence should be addressed to; jjli@uw.edu. Electronic Supplementary Information (ESI) available
[a] Those authors contributed equally to the work.

# INTRODUCTION

Solid state thermoelectric conversion is promising for recovering tremendous waste heat produced by human society and for enabling more effective thermal management. However, the high thermoelectric conversion efficiency, governed by the dimensionless figure of merit *ZT*, requires simultaneously high electrical conductivity and Seebeck coefficient, yet low thermal conductivity, which are rather difficult to obtain in a single-phase material [1-5]. One of the primary mechanisms for improving thermoelectric performance in a material is to reduce lattice thermal conductivity through scattering of phonons by structural heterogeneity such as defects, interfaces, and impurities. In the last two decades, the nanostructuring paradigm has been successfully applied to a wide range of thermoelectric materials, resulting in significant reduction in thermal conductivity and superior thermoelectric performance [6-12]. For example, nanocrystalline $(Bi,Sb)_2Te_3$ with maximum *ZT* values of 1.4-1.6 were reported [7, 13-15], and the improvement was largely attributed to the low lattice thermal conductivity resulting from intensified phonon scattering by nanostructures and defects [16]. In addition, by combining all-scale hierarchical architectural microstructures, including atomic defects, endotaxial nanoprecipitates, and mesoscale grains [17], a wide range of heat carrying phonons can be strongly scattered [9], resulting in record high *ZTs* in PbTe [17]. Similar reduction in thermal conductivity has also been reported in many nanocomposites, for example systems consisting of half-Heusler/full-Heusler and $In_xCe_yCo_4Sb_{12}$/InSb [18].

These advances in thermoelectric materials highlight the importance of nanostructuring in enhancing thermoelectric properties, yet quite surprisingly, such improvements have been largely accomplished with no direct investigation of the local thermoelectric properties in nanostructured materials, even though local electric current can be mapped with high spatial resolution. Indeed, there still lacks an effective method that directly links the macroscopic thermoelectric performance to the local microstructures and properties. While the microstructural heterogeneity can be mapped with *atomic* resolution in terms of chemical and phase compositions, it reveals little about local transport behavior. Traditionally, the thermoelectric properties are only measured at the macroscopic scale, averaged over various microstructural features, and it is rather difficult to know exactly and directly *which* material constituent contributes to *what* in the local thermal transport processes. Therefore, high spatial resolution is critically needed in the thermal analysis of nanostructured materials in general, and nanostructured thermoelectrics in particular.

In the last decade, time domain thermoreflectance (TDTR) has emerged as a powerful tool for thermal transport property measurements [19, 20], though its spatial resolution is limited to hundreds of nanometers [21-24], making the detailed mapping of thermal transport properties in nanostructured materials challenging. Various scanning thermal microscopy (SThM) techniques have been developed based on temperature-sensitive phenomena in the sample, promising potentially higher spatial resolutions [25-36], though they have rarely provided even a qualitative thermal mapping that correlates with the microstructures in a heterogeneous material, and strong crosstalk between thermal imaging and surface topography is often observed. Such direct correlation, especially in quantitative manner, is highly desirable for understanding as well as further design and optimization of high performance thermoelectric materials, which we seek to accomplish via tightly coupled SThM experiments and finite element simulations utilizing a resistive heating thermal probe.

In this work, we show that local thermal conductivity can be mapped quantitatively with good accuracy, nanometer resolution and *one-to-one* correspondence to the microstructure, using filled skutterudite as a model system. Yb-filled $CoSb_3$, with its maximum *ZT* up to 1.5, is one of the most promising thermoelectric materials for applications in the intermediate temperature range, since guest filling in the structural nanovoids acts to control the carrier concentration, while significantly suppress the propagation of heat-carrying phonons [37-39]. However, impurity phases such as $YbSb_2$, $Yb_2O_3$, and $CoSb_2$ are commonly observed due to the filling fraction limit of approximately 0.3 in $Yb_xCo_4Sb_{12}$, low formation energy of Yb oxide, and complexity in the Yb-Co-Sb phase diagram, and these impurity phases could exert significant influence on thermoelectric properties [40]. It thus provides us an ideal model system to study its local thermal and electric conductivities.

**RESULTS**

*Three-phase Microstructure*

The sample with a stoichiometry of $Yb_{0.7}Co_4Sb_{12}$ was prepared by a conventional induction melting-vacuum melting-quenching-annealing-sintering method, resulting in a three-phase microstructure as shown in Fig. 1. A back-scattered electron (BSE) image of the sample clearly reveals three phases as marked (Fig. 1(a)), with impurity phases (2 and 3) embedded in the matrix phase (1). It is anticipated that the excessively added Yb reacts with Sb to form the $YbSb_2$ phase, and the resultant Sb-deficiency leads to the formation of CoSb$_2$. Such scenarios are indeed confirmed by the elemental mappings of Yb, Co, Sb, and O (Fig. 1 (b-e)) as well as elemental ratios in each phases determined from the energy-dispersive X-ray spectroscopy (EDS) (Fig. 1(f) and Fig. S1 in Supplementary Information (SI)), suggesting that the matrix phase (1) is filled skutterudite Yb$_{0.3}$Co$_4$Sb$_{12}$, while the impurity phases (2 and 3) are CoSb$_2$ and surface oxidized YbSb$_2$, respectively. This analysis of phase composition is also confirmed by X-ray diffraction shown in Fig. S2 in SI.

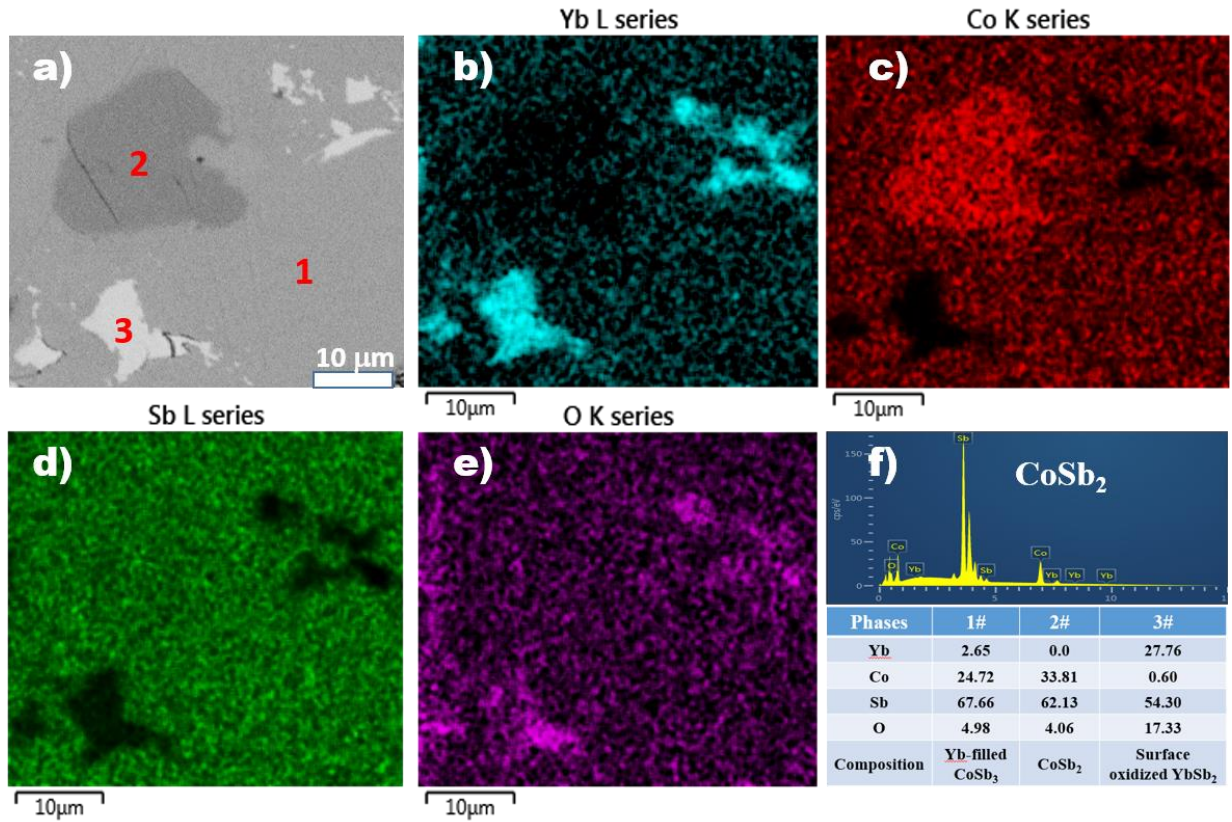

**Fig. 1** Composition analyses of three-phase microstructure in a Yb$_{0.7}$Co$_4$Sb$_{12}$; (a) a typical BSE image, wherein the three phases are labeled as 1, 2 and 3; (b)-(e) corresponding elemental mappings of Yb, Co, Sb, and O for the area shown in (a); (f) a typical EDS spectrum of phase 1, and elemental ratios of three different phases labeled in (a).

*Scanning Thermal Microscopy*

The microstructural analysis in Fig. 1 is powerful in mapping local chemical composition and phase structure of a material, yet it reveals nothing about the local thermoelectric properties. Traditionally, such properties are measured at the macroscopic scale, and their local variations, if any, can only be deduced indirectly from the macroscopic measurement or from computational analyses. If one can correlate local thermal properties directly with the microstructural features, then the effect of structural heterogeneity on the macroscopic thermoelectric conversion can be better understood and optimized. We seek to accomplish this via a scanning thermal probe [29, 41, 42], as shown in Fig. 2(a). It has a micro-fabricated solid-state resistive heater at the end of the cantilever [31, 43-46], which forms one branch of the Wheatstone bridge circuit that allows precise measurement of its electrical resistance. When the probe scan phases with lower (Fig. 2(b)) and higher (Fig. 2(c)) thermal conductivities, it will have lower and higher temperature drops, respectively, resulting in different probe resistances that can be measured accurately from the imbalanced Wheatstone bridge voltage. This enables us to image local thermal response quantitatively. Indeed, as shown in Fig. 2(d), there is a linear relationship between probe resistance and temperature around the SThM operation temperature, calibrated by passively probing a hot-plate stage with well-defined temperatures:

$$R(T) = R_0\big(1 + \alpha(T - T_0)\big), \tag{1}$$

where $R_0$ is the resistance of the thermal probe at room-temperature reference $T_0 = 293.15$ K, and $\alpha$ is the temperature coefficient of resistance (TCR) measured to be 0.8/K between 350 and 450 K (Fig. 2(d)), the temperature range relevant or our subsequent experiments. As such, the thermal probe not only functions as a heater, but also as a local temperature sensor via the resistance measurement, making it possible to measure the local thermal properties of the sample quantitatively [25, 47].

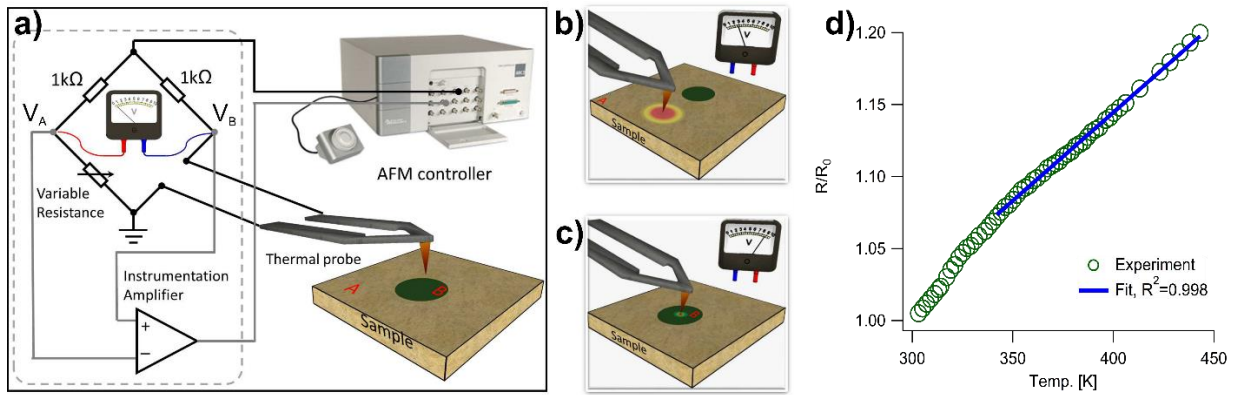

**Fig. 2** The schematics of SThM setup; (a) resistive heating thermal probe in a balanced Wheatstone bridge circuit ($V_A - V_B = 0$) before touching the sample; the thermal probe scans phases with (b) lower and (c) higher thermal conductivities, resulting in different probe temperatures that can be measured via imbalanced Wheatstone bridge for imaging; (d) linear correlation between temperature and resistance of the thermal probe calibrated by a hot plate with known temperatures.

In order to precisely measure the change in resistance induced by heat transfer to areas with different thermal conductivities, the Wheatstone bridge is balanced by adjusting the variable resistor before contacting the sample, and the bridge voltage is amplified with a 10× gain using a differential amplifier for enhanced sensitivity. After it touches the sample, heat transfers from the probe to the sample, resulting in drops of its temperature and resistance, and thus a voltage drop between nodes A and B of the Wheatstone bridge. Lower sample thermal conductivity has lower heat loss and thus lower drop in resistance, corresponding to lower bridge voltage difference (Fig. 2(b)). Hence higher voltage difference indicates a higher sample thermal conductivity (Fig.2(c)), making it possible to image local thermal conductivity variation based on the bridge voltage.

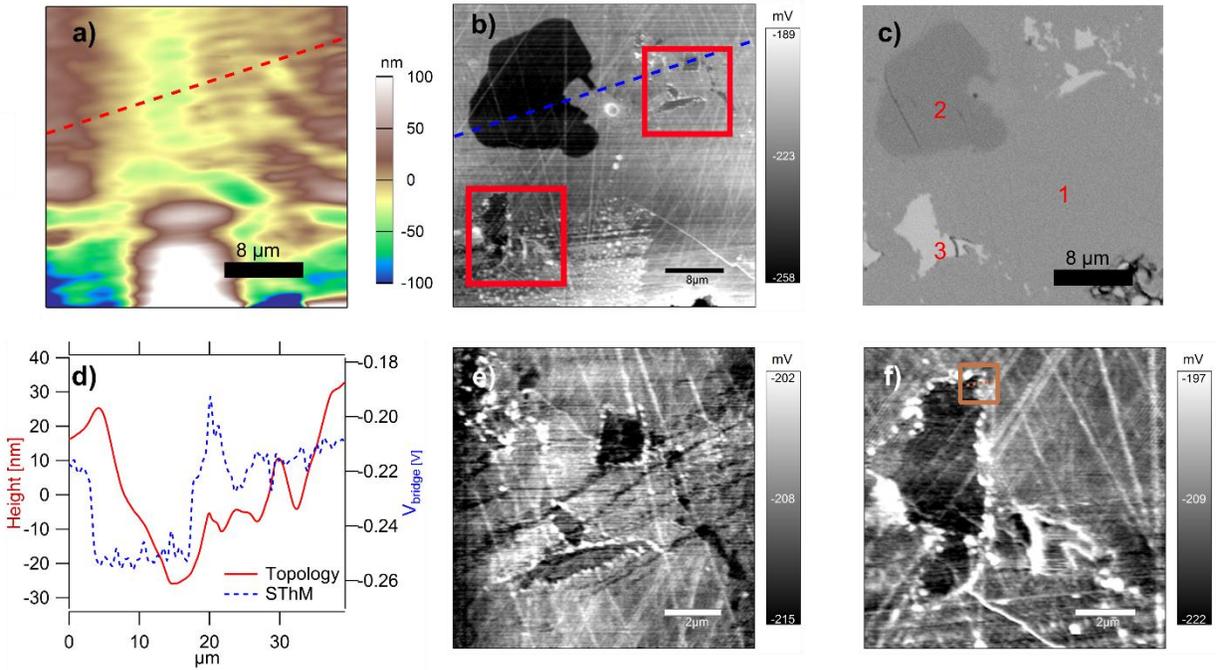

**Fig. 3** SThM mapping of $Yb_{0.7}Co_4Sb_{12}$; (a) topography; (b) distribution of Wheatstone bridge voltage with 10× amplification; (c) BSE image; (d) comparison of line scans in topography and voltage mappings; (e) and (f) higher resolution voltage mapping in smaller areas.

With a marker made near the area of the BSE image (Fig. S3), we were able to locate the same area in our SThM studies. The topography mapping in Fig. 3(a) obtained from the contact mode reveals relatively flat surface without any correlation with the microstructure shown in BSE image of Fig. 1. However, simultaneous to the topography scan, the mapping of Wheatstone bridge voltage difference in Fig. 3(b) clearly reveals three different contrasts that correlate well with the BSE microstructure, repeated here in Fig. 3(c) for a direct comparison. Due to the drop in the probe temperature, and thus the resistance, the Wheatstone bridge voltage drops to negative. The matrix phase (1) has the smallest voltage drop, while the impurity phase (2) has the largest. Such contrast is induced by the difference in thermal conductivity, not from the topography variation, as evident by the line scan comparison in Fig. 3(d). When a phase interface is crossed, sharp change in voltage difference is observed, while topography is relatively flat. On the other hand, within an individual phase, voltage difference is roughly constant, while topography variation is observed. The voltage difference mapped thus reflects the variation in thermal conductivity with no crosstalk to topography, wherein the matrix phase has the lowest thermal conductivity, while the impurity phase (2) has the highest. Thus, this is a direct characterization of local thermal conductivity that illustrates the role of impurity phases in thermal transport. Higher resolution scans of two boxes marked in Fig. 3(b) are shown in Figs. 3(e,f), demonstrating even higher sensitivity and spatial resolution. It is worth noting that although YbSb$_2$ (phase 3) is supposed to have the highest thermal conductivity, it appears that the significant surface oxidation reduces its thermal conductivity, resulting in intermediate voltage drops as mapped.

*Finite Element Simulation*

In order to interpret and analyze the SThM data quantitatively, finite element model (FEM) implemented in the COMSOL Multiphysics package was developed to study heat transfer among the thermal probe, the sample, and surrounding air, as shown in Fig. S4. Two dominant physical processes were considered, one is Joule heating in the resistive heater of thermal probe, and the other is heat conduction in the thermal probe, sample, and surrounding air. The quasi-steady problem was solved using a stationary solver based on the conduction equation

$$-\nabla \cdot (\kappa \nabla T) = Q \,, \tag{2}$$

where $\kappa$ is the thermal properties of each domain listed in Table S1, and $Q$ is the heat source, set to be zero in all domains except in the resistive heater part of the thermal probe under an input voltage $V_0$,

$$Q = \frac{V_0^2}{R(T)}. \tag{3}$$

It is important to recognize that the temperature of the thermal probe and its resistance are intimately coupled, resulting in a nonlinear governing equation to solve. In the simulations, the initial value of temperature is set as $T_0 = 293.15$ K for all domains, and the external boundaries of air-box and the sample (far from the thermal probe) are set to have ambient temperature $T = T_0$ as well. Across interfaces between different domains, temperature and heat flux density are assumed to be continuous, except on the tip-sample contact wherein a thermal contact resistance of $1.0\times10^8$ K/W was defined, which was measured using similar type of SThM probe in a previous study [48]. The presence of the contact resistance results in significant temperature drop at the tip-sample junction but rather small temperature increase in resistive heater that is more relevant for the measurement (Fig. S5). More importantly, the temperature of resistive heater is sensitive to the change in thermal conductivity of the sample, but insensitive to the contact resistance (Fig. S6), as the total heat loss due to the contact resistance is less than 1 µW while the total heat conductance through tip-sample junction is in order of 100 µW. We also point out that the radiative and convective heat losses, as shown in Figs. S7 and S8, are negligible in comparison to the heat conduction, and thus are not considered. More detailed analysis can be found in SI.

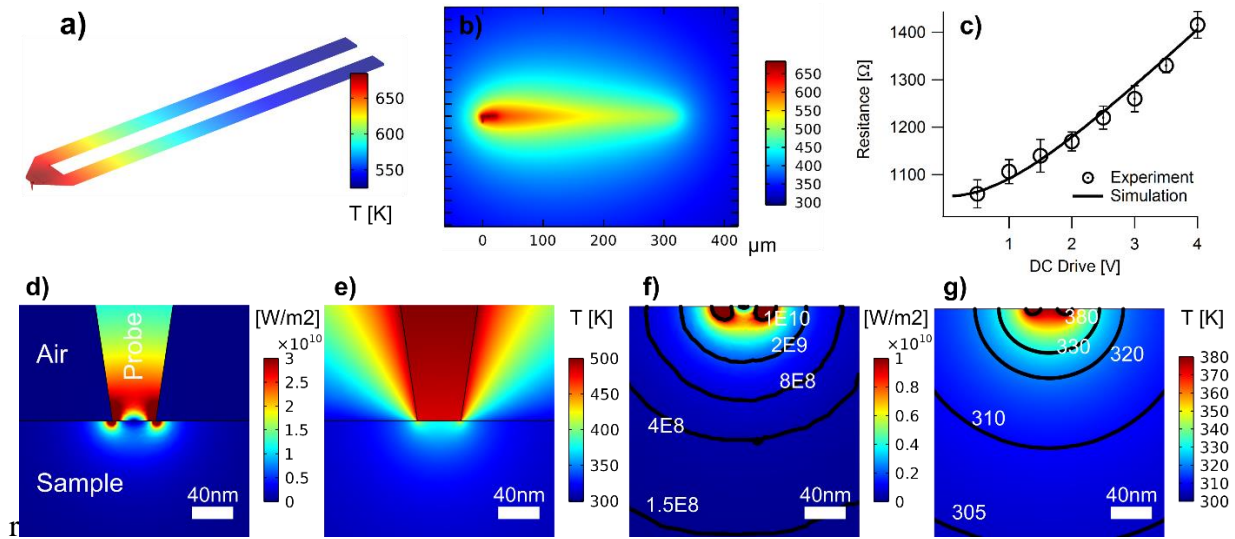

**Fig. 4** FEM simulation of the SThM experiment; (a-c) when the probe is held in air: (a) 3D and (b) cross-sectional temperature distribution of thermal probe under 3.5V; (c) measured probe resistance as a function of DC drive voltage in comparison with simulation; (d-g) when the probe is in contact with a homogeneous sample having $\kappa = 5\text{W}/(\text{m}.\text{K})$, and the contact resistance is taken to be $1.0\times10^8$ K/W; cross-sectional distribution of (d) heat flux density and (e) temperature

on the tip-sample junction; and overlaid contour on the distribution of (f) heat flux distribution and (g) temperature in the sample underneath the probe.

We first examine the thermal probe in air, far away from the sample. Under an input voltage of 3.5V, the temperature distribution is shown in Fig. 4(a-b), indicating a substantial temperature rise in the thermal probe up to 683K. The resistance of the probe is expected to rise with the increased input voltage in a nonlinear manner, caused by higher temperature induced by the heating voltage. This is confirmed in Fig. 4(c), wherein good agreement between experimental measurement and FEM simulation is observed, giving us confidence that the simulation does capture the physical processes in SThM experiment accurately. We then in our simulation bring the thermal probe in contact with a sample having a thermal conductivity of $\kappa = 5\text{W}/(\text{m.K})$, and the increased heat conduction through probe-sample junction results in a drop in the probe temperature, and thus a drop in its resistance. Analysis on the overall heat flux and temperature distributions suggest that only a few percentage of the generated heat passes to the sample through the contact via heat conduction. Nevertheless, due to the small contact area of probe-sample junction, the heat flux density is substantial, and the resulting temperature drop in the thermal probe is significant, around 190 K. These are evident in the distributions of heat flux density and temperature in the probe-sample junction shown in Fig. 4(d-e). Note the significant temperature drop observed at the contact, yet the problem can be treated a quasi-static, and thus the temperature drop does not significantly influence the temperature of the resistive heater, as compared in Fig. S5. Further zoom-in on the sample in Fig. 4(f-g) shows that the radius of the sample thermal volume affected by the probe is less than 100nm, within which around 90% of temperature variation in the sample occur. This confirms the nanoscale resolution of our SThM technique.

*Quantitative Mappings of Local Properties*

Are we capable of experimentally distinguishing materials with different thermal conductivities then? To answer this question, both FEM simulations and SThM experiments were carried out on a dozen or so samples with nominal thermal conductivities ranging from 0.66 to 80.8 W/mK, spanning two orders of magnitude, as listed in Table S2. Under a constant 3.5V input to the thermal probe, the drop in the resistances when the probe contacts the samples was predicted by FEM simulations and measured by SThM experiments, as shown in Fig. 5(a). The point-wise experiments were repeated 5 times on each calibration sample in different areas, and the mean and standard deviation were obtained with the SThM contact force kept constant throughout all the experiments. In the simulation, a constant contact resistance of $1.0\times10^8$ K/W is assumed, as explained earlier, and the results suggest that the thermal contact resistance is only significant for samples with higher thermal conductivity, consistent with what we observed in Fig. S6. Good agreement between simulations and experiments observed in Fig. 5(a) suggests that the thermal conductivity of the sample and the resistance drop can be quantitatively correlated. Indeed, for the three-phase microstructure mapped in Fig. 3, the distribution of resistance change, shown here in Fig. 5(b), can be directly converted into a mapping of thermal conductivity shown in Fig. 5(c) based on this correlation. The resulted thermal conductivities in the three phases are summarized in Table 1, and good agreements with nominal values reported in literature are observed [39, 49, 50], validating quantitative SThM mapping. Of particular interests is the variation of thermal response at an interface between two phases, for which we investigated an interface marked by the dashed line in the box in Fig. 3(f), passing from phase 3 to phase 1. The variation of expected resistance change when the probe scans across the interface is simulated by FEM and compared with experiments, and again good agreement is observed, as shown in Fig. 5(d), with quantitative difference much less than 1%. FEM simulations were carried out under nominal thermal conductivity distributions with a sharp interface, as indicated by the blue line in Fig. 5(d), while the transition length of thermal response variation is in the order of 100 nm for both experiment and simulation. Note that we have not considered the interfacial resistance in our analysis.

**Table 1** Comparison of thermal conductivities measured from SThM experiment and reported in literature at 300K

| $\kappa$ [W/(m.K)] | $Yb_{0.3}Co_4Sb_{12}$[†] | $CoSb_2$[#] | $YbSb_2$ | $Yb_2O_3$[*] |
|---|---|---|---|---|
| Phase | 1 | 2 | 3 | 3 |
| Measured | 4.63 | 11.71 | - | 9.52 |
| Reported | 3.2 | 11.8 | 15.0 | - |
| Remarks | Polycrystal | Single-crystal | Polycrystal | Surface oxidized |

† Ref. [39],   # Ref. [49],   *Ref. [50]

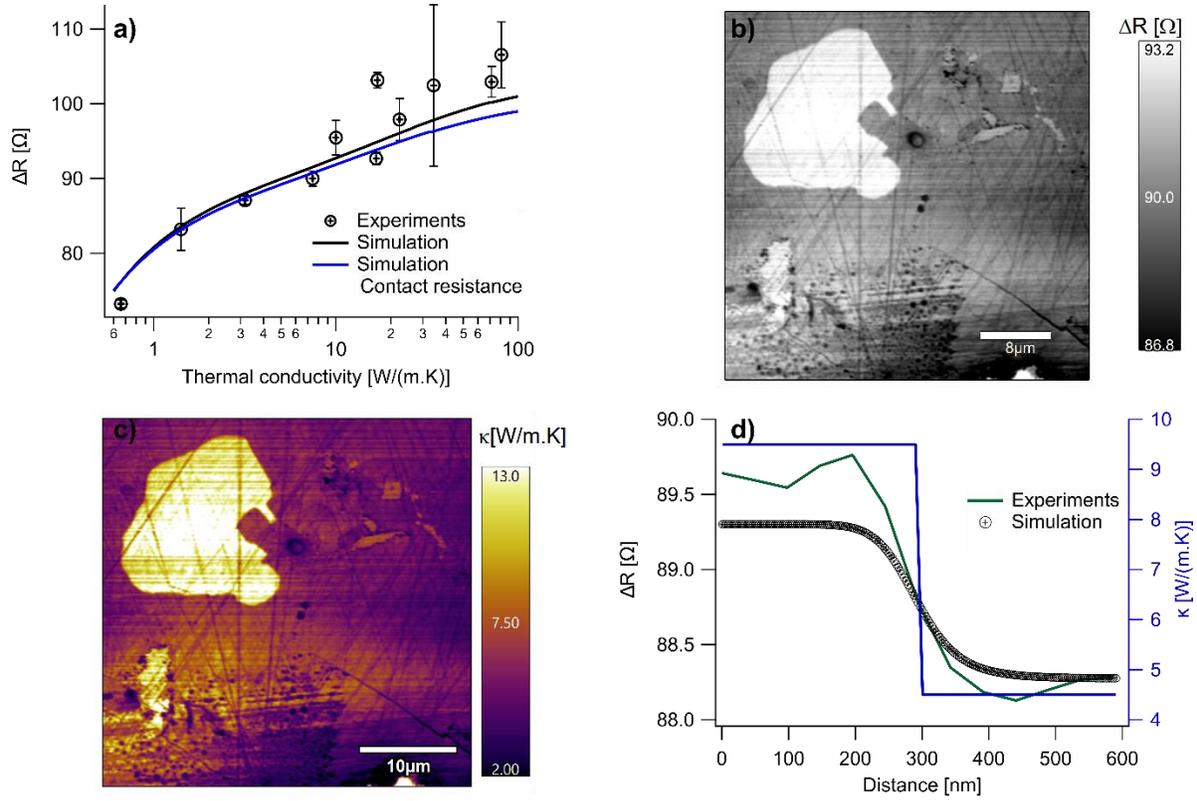

**Fig. 5** Quantitative mapping of thermal conductivities; (a) changes in the probe resistance induced by samples with different thermal conductivities; and mappings of (b) resistance change and (c) corresponding thermal conductivities in $Yb_{0.7}Co_4Sb_{12}$; (d) line scan of resistance change across an interface between phase 1 and phase 3.

We also examined the electrical conductivity of the sample in the same area using a conductive AFM (cAFM), as shown in Fig. 6, where good correlation between current mapping and phase distribution is again evident. The current mapping in Fig. 6(b) suggests that the electrical conductivity in the matrix phase 1 and secondary phase 2 is relatively high, while that of phase 3 is relatively low due to surface oxidation that we discussed earlier. As shown in Table S3, phases 1 and 2 have comparable electrical conductivities of $2.3 \times 10^5$ S/m and $3.0 \times 10^5$ S/m, respectively, consistent with the slight contrast in Fig. 6(b). For phase 3, the oxidation of $YbSb_2$ introduces an insulating layer on the surface, resulting in very low electrical conductivity. The I-V curves measured in each phase shown in Fig. 6(c) also confirm this observation. Obviously, these impurity phases with higher thermal conductivity but lower electrical conductivity exert detrimental influence on the thermoelectric properties of skutterudites, and thus are not desirable. Indeed, the thermoelectric properties measured at the macroscopic scale, as shown in Fig. S9, confirmed this analysis, that its figure of merit *ZT* is not optimal due to these impurity phases. As such, in order to exert significant scattering on phonons while negligible influence on electrons, it is critical to control precisely the stoichiometry and phase composition, as well as the size and morphology of impurity phases in filled skutterudites. In fact, this is important for nanostructured thermoelectrics in general.

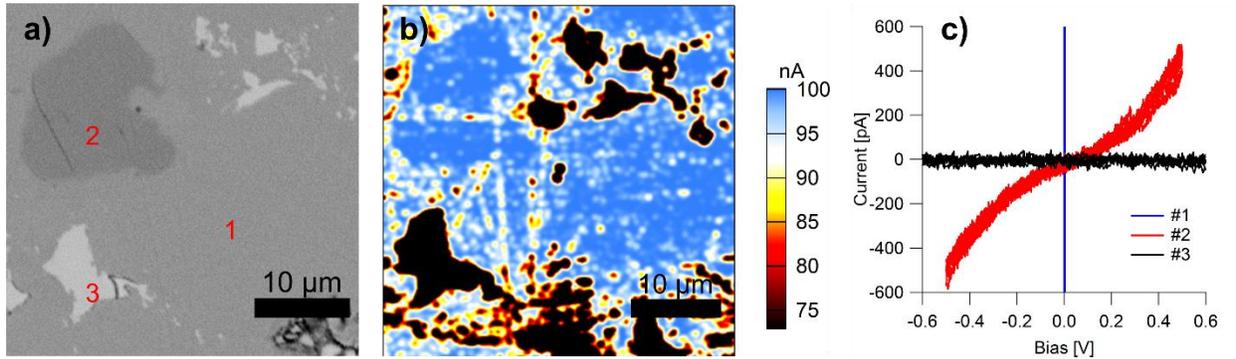

**Fig. 6** Mapping of local electric conductivity; (a) BSE image; (b) mapping of current; (c) IV curves measured in three phases using cAFM.

**DISCUSSIONS**

It is rather remarkable that small contrasts in local thermal conductivity can be accurately mapped with nanoscale resolution via SThM as demonstrated. To help us understand this better, we compare the FEM-simulated heat transfer processes involving 3 representative phases: matrix skutterudite $Yb_{0.3}Co_4Sb_{12}$ (1), impurity phase $CoSb_2$ (2), and surface oxidized phase $Yb_2O_3$ (3), using as the input their nominal thermal conductivities of 4.5, 11, and 9.5 W/(m.K), respectively. The relevant thermal transport parameters computed are summarized in Table 2. It is interesting to note that although less than 2% of the total heat generated transfers through the tip-sample junction for each of the phase considered, the process, i.e. the *change* in heat transfer, is still dominated by the conductivity of the sample, and thus the heat flux at the tip-sample junction changes significantly for samples with different thermal conductivities, resulting in different temperature of the contact and thus different probe temperature and resistance that can be precisely measured using a Wheatstone bridge. This confirms the feasibility of the proposed technique and validates the experimental data. Further improvement can be achieved by carrying out SThM in vacuum [51], which would enhance sensitivity and resolution, though our study show that it is not absolutely necessary.

**Table 2** Numerical heat transfer properties at the contact of thermal probe with different phases of the sample

| $\kappa$ [W/(m.K)] | Source [mW] | Contact heat conduction/source [%] | Contact heat flux [GW/m²] | flux air to sample [kW/m²] | Contact temp. [K] | Probe temp [K] | $\Delta R$ [Ω] |
|---|---|---|---|---|---|---|---|
| 4.5 | 9.968 | 0.9805 | -4.861 | -22.352 | 442.74 | 525.64 | 88.81 |
| 11 | 9.994 | 1.208 | -6.105 | -22.069 | 410.58 | 521.44 | 92.38 |
| 9.5 | 9.990 | 1.229 | -6.004 | -22.127 | 416.33 | 522.12 | 91.72 |

It is also important to examine the transient behavior of heat transfer during the SThM scanning, which is shown in Fig. S10 obtained from FEM simulations. It takes less than 1.2 milliseconds for the probe temperature to reach its steady-state value, and this is much less than the pixel time of SThM (3.1mSec for 0.5Hz line scan), ensuring a quasi-steady-state condition during the SThM scanning. This also imposes an upper limit on scanning rate at 1.3Hz per line. Finally, we note that the spatial resolution of this technique can be further improved using a dynamic approach. Under a quasi-steady state condition, the affected thermal volume in the sample is substantial, with radius on the order of 100 nm. This has been confirmed by the temperature distribution in the sample underneath of the probe (Fig. 4(g)), as well as the line scan of resistance change across an interface (Fig. 5(d)). If a dynamic probing is adopted instead, wherein the heating voltage of the probe is frequency-modulated, and the corresponding harmonic response is probed, much higher spatial resolution on the order of 10 nm is expected. This will provide us a powerful tool to study the effect of structural heterogeneity such as defects, interfaces, and impurities in thermoelectric materials, which can be used to guide the design and optimization of thermoelectric materials with enhanced performance, especially when the thermoelectric coefficient can also be mapped in addition to thermal conductivity and electric current.

**CONCLUSION**

In this work, local thermal conductivities of a three-phase filled skutterudite were mapped quantitatively with good accuracy, nanometer resolution, and one-to-one correspondence with microstructure. Quantitative mapping was accomplished via a SThM using resistive heating thermal probe complemented by FEM simulations, enabling us to distinguish thermal conductivities spanning two orders of magnitude, yet resolving thermal variation across a phase interface with small thermal conductivity contrast. The technique developed thus provides a powerful tool to correlate local thermal conductivities, microstructures, and macroscopic properties for nanostructured materials in general, and thermoelectric materials in particular, which can be used to guide the design and optimization of thermoelectric materials with enhanced performance.

**METHODS**

*Sample preparation*

The sample with a stoichiometry of $Yb_{0.7}Co_4Sb_{12}$ was prepared by a conventional induction melting-vacuum melting-quenching-annealing-sintering method. High purity Co powders (99.995%, Alfa Aesar), Sb shots (99.9999%, Alfa Aesar), and Yb chunks (99.95%, Alfa Aesar) were used as the starting materials. Co powders were first purified and melted into small shots with sizes of 1-5 mm by arc melting (SA-200, MRF Inc., USA), then loaded into a BN crucible together with Sb shots for induction melting (at 2000 ºC for 30 s under an Ar atmosphere, EQ-SP-25VIM, MTI Corporation, USA). The obtained ingot was subsequently crushed and loaded into a carbon-coated quartz tube with appropriate amounts of Yb and Sb in an argon-filled glove-box (Lab Star, Mbraun, Germany) and vacuum sealed ($10^{-3}$ torr). Subsequently, the tube with raw materials was placed into a box furnace, heated to 1000 ºC in 5 h, soaked for 24 h, and then rapidly quenched in ice water. The obtained ingot was ultrasonically cleaned and vacuum-sealed in a quartz tube, then annealed in a box furnace at 750 ºC for 168 h. After annealing, the ingot was crushed, hand grounded into fine powders and sintered into a bulk material using the spark plasma sintering (SPS-211Lx, Dr. Sinter, Japan) at 680 ºC and 50 MPa for 5 min.

*Structure and property characterization*

The phase composition was determined by the powder X-ray diffraction (XRD, Bruker D8 Focus X-ray diffraction, Germany) using the Cu K$_\alpha$ radiation ($\lambda$= 1.5406 Å). The BSE images were obtained in a TM3000 electron microscope (Hitachi, Japan). The chemical composition and elemental mapping were determined by a field emission scanning electron microscope equipped with EDS (FESEM/EDS, FEI Sirion, Japan). The electrical conductivity and Seebeck coefficient were simultaneously measured via commercial equipment (ZEM-3, Ulvac Riko, Inc., Japan) under a low-pressure helium atmosphere. Thermal conductivity was calculated from the product of the measured thermal diffusivity, specific heat, and density. Thermal diffusivity was measured by a laser flash method (Netzsch LFA-457, Germany), and specific heat was measured by a differential scanning calorimetry method (DSC) using sapphire as the reference (Netzsch 404F1, Germany). The measurement temperature ranges from 300 K to 850 K.

*Scanning thermal microscopy*

The scanning thermal microscopy was performed with an Asylum Research MFP-3D atomic force microscope (AFM) using a thermal probe with a spring constant of 0.2-0.5 N/m (AN2-300, Anasys Instruments) in contact mode with contact force around 110nN and scan rate of 0.5 Hz per line. The thermal probe is similar to any silicon AFM probe in geometry with an integrated heater at the end. A small region (7.5 $\mu m \times 15\ \mu m$) close to the cantilever tip with light phosphorus doping acts as a solid-state heater. The heater region connects through two heavily doped branches acting as the electrical leads. The electrical resistivity of the probe varies between 600 Ω and 2 $k\Omega$ for different probes and different drive voltages.

*Conductive AFM*

The variations of electrical conductivity of the sample were characterized using ORCA, a conductive AFM (cAFM) module developed by Asylum Research. All measurements and images were obtained using AFM contact mode with a metallic Pt/lr coated probe having a force constant of 2.8 N/m (PPP-EFM, Nanosensors). Two different modes were used: imaging mode to map the variation of electrical conductivity, and spectroscopic mode to measure the I-V characteristics at points of interest. In the imaging mode, 2V DC bias was applied to the sample substrate and the necessary current for virtually keeping the tip ground was used to image the electrical conductivity. In the spectroscopic mode under stationary probe, a sweeping DC bias (-0.6 V to 0.6 V) was applied to the sample while measuring the current that kept the conductive tip ground.

**ACKNOWLEDGEMENTS**

We acknowledge National Key Research and Development Program of China (2016YFA0201001), National Natural Science Foundation of China (11627801 and 11472236), and US National Science Foundation (CBET-1435968). This material is based in part upon work supported by the State of Washington through the University of Washington Clean Energy Institute.

**REFERENCES**

1. Snyder GJ, Toberer ES; Complex thermoelectric materials. *Nat Mater* 2008;**7**(2):105-14.
2. Venkatasubramanian R, Siivola E, Colpitts T, et al.; Thin-film thermoelectric devices with high room-temperature figures of merit. *Nature* 2001;**413**(6856):597-602.
3. Tritt TM, Subramanian MA; Thermoelectric materials, phenomena, and applications: A bird's eye view. *Mrs Bulletin* 2006;**31**(3):188-194.
4. Zebarjadi M, Esfarjani K, Dresselhaus MS, et al.; Perspectives on thermoelectrics: from fundamentals to device applications. *Energy & Environmental Science* 2012;**5**(1):5147-5162.
5. Zhao LD, Dravid VP, Kanatzidis MG; The panoscopic approach to high performance thermoelectrics. *Energy & Environmental Science* 2014;**7**(1):251-268.
6. Dresselhaus MS, Chen G, Tang MY, et al.; New Directions for Low‐Dimensional Thermoelectric Materials. *Adv Mater* 2007;**19**(8):1043-1053.
7. Poudel B, Hao Q, Ma Y, et al.; High-thermoelectric performance of nanostructured bismuth antimony telluride bulk alloys. *Science* 2008;**320**(5876):634-8.
8. Minnich AJ, Dresselhaus MS, Ren ZF, et al.; Bulk nanostructured thermoelectric materials: current research and future prospects. *Energy & Environmental Science* 2009;**2**(5):466-479.
9. Vineis CJ, Shakouri A, Majumdar A, et al.; Nanostructured thermoelectrics: big efficiency gains from small features. *Adv Mater* 2010;**22**(36):3970-80.
10. Joshi G, Lee H, Lan Y, et al.; Enhanced thermoelectric figure-of-merit in nanostructured p-type silicon germanium bulk alloys. *Nano Lett* 2008;**8**(12):4670-4.
11. Pei YZ, Heinz NA, LaLonde A, et al.; Combination of large nanostructures and complex band structure for high performance thermoelectric lead telluride. *Energy & Environmental Science* 2011;**4**(9):3640-3645.
12. Vaqueiro P, Powell AV; Recent developments in nanostructured materials for high-performance thermoelectrics. *Journal of Materials Chemistry* 2010;**20**(43):9577-9584.
13. Xie W, He J, Kang HJ, et al.; Identifying the specific nanostructures responsible for the high thermoelectric performance of (Bi,Sb)2Te3 nanocomposites. *Nano Lett* 2010;**10**(9):3283-9.
14. Cao YQ, Zhao XB, Zhu TJ, et al.; Syntheses and thermoelectric properties of Bi2Te3/Sb2Te3 bulk nanocomposites with laminated nanostructure. *Applied Physics Letters* 2008;**92**(14):3106.
15. Guin SN, Chatterjee A, Negi DS, et al.; High thermoelectric performance in tellurium free p-type AgSbSe 2. *Energy & Environmental Science* 2013;**6**(9):2603-2608.
16. Poudeu PFP, D'Angelo J, Kong H, et al.; Nanostructures versus Solid Solutions: Low Lattice Thermal Conductivity and Enhanced Thermoelectric Figure of Merit in Pb9. 6Sb0. 2Te10-x Se x Bulk Materials. *Journal of the American Chemical Society* 2006;**128**(44):14347-14355.
17. Biswas K, He J, Blum ID, et al.; High-performance bulk thermoelectrics with all-scale hierarchical architectures. *Nature* 2012;**489**(7416):414-8.
18. Li H, Tang XF, Zhang QJ, et al.; High performance InxCeyCo4Sb12 thermoelectric materials with in situ forming nanostructured InSb phase. *Applied Physics Letters* 2009;**94**(10):102114.
19. Paddock CA, Eesley GL; Transient Thermoreflectance from Thin Metal-Films. *Journal of Applied Physics* 1986;**60**(1):285-290.

20. Cahill DG, Goodson KE, Majumdar A; Thermometry and thermal transport in micro/nanoscale solid-state devices and structures. *Journal of Heat Transfer-Transactions of the Asme* 2002;**124**(2):223-241.
21. Christofferson J, Shakouri A; Thermoreflectance based thermal microscope. *Review of Scientific Instruments* 2005;**76**(2):24903-24903.
22. Huxtable S, Cahill DG, Fauconnier V, et al.; Thermal conductivity imaging at micrometre-scale resolution for combinatorial studies of materials. *Nat Mater* 2004;**3**(5):298-301.
23. Cahill DG; Analysis of heat flow in layered structures for time-domain thermoreflectance. *Review of Scientific Instruments* 2004;**75**(12):5119-5122.
24. Hatori K, Taketoshi N, Baba T, et al.; Thermoreflectance technique to measure thermal effusivity distribution with high spatial resolution. *Review of Scientific Instruments* 2005;**76**(11):114901.
25. Nonnenmacher M, Wickramasinghe HK; Scanning Probe Microscopy of Thermal-Conductivity and Subsurface Properties. *Applied Physics Letters* 1992;**61**(2):168-170.
26. Majumdar A, Carrejo JP, Lai J; Thermal Imaging Using the Atomic Force Microscope. *Applied Physics Letters* 1993;**62**(20):2501-2503.
27. Luo K, Shi Z, Varesi J, et al.; Sensor nanofabrication, performance, and conduction mechanisms in scanning thermal microscopy. *Journal of Vacuum Science & Technology B* 1997;**15**(2):349-360.
28. Mills G, Zhou H, Midha A, et al.; Scanning thermal microscopy using batch fabricated thermocouple probes. *Applied Physics Letters* 1998;**72**(22):2900-2902.
29. Majumdar A; Scanning thermal microscopy. *Annual Review of Materials Science* 1999;**29**(1):505-585.
30. Fiege GBM, Altes A, Heiderhoff B, et al.; Quantitative thermal conductivity measurements with nanometre resolution. *Journal of Physics D-Applied Physics* 1999;**32**(5):L13-L17.
31. King WP, Kenny TW, Goodson KE, et al.; Design of atomic force microscope cantilevers for combined thermomechanical writing and thermal reading in array operation. *Journal of Microelectromechanical Systems* 2002;**11**(6):765-774.
32. Lyeo HK, Khajetoorians AA, Shi L, et al.; Profiling the thermoelectric power of semiconductor junctions with nanometer resolution. *Science* 2004;**303**(5659):816-8.
33. Kim K, Jeong W, Lee W, et al.; Ultra-high vacuum scanning thermal microscopy for nanometer resolution quantitative thermometry. *ACS Nano* 2012;**6**(5):4248-57.
34. Rojo MM, Martin J, Grauby S, et al.; Decrease in thermal conductivity in polymeric P3HT nanowires by size-reduction induced by crystal orientation: new approaches towards thermal transport engineering of organic materials. *Nanoscale* 2014;**6**(14):7858-65.
35. Pumarol ME, Rosamond MC, Tovee P, et al.; Direct nanoscale imaging of ballistic and diffusive thermal transport in graphene nanostructures. *Nano Lett* 2012;**12**(6):2906-11.
36. Grauby S, Puyoo E, Rampnoux JM, et al.; Si and SiGe nanowires: fabrication process and thermal conductivity measurement by 3ω-scanning thermal microscopy. *The Journal of Physical Chemistry C* 2013;**117**(17):9025-9034.
37. Zhao XY, Shi X, Chen LD, et al.; Synthesis of YbyCo4Sb12/Yb2O3 composites and their thermoelectric properties. *Applied Physics Letters* 2006;**89**(9):92121-92121.
38. Dahal T, Jie Q, Joshi G, et al.; Thermoelectric property enhancement in Yb-doped n-type skutterudites YbxCo 4 Sb 12. *Acta Materialia* 2014;**75**:316-321.
39. Wang SY, Salvador JR, Yang J, et al.; High-performance n-type YbxCo4Sb12: from partially filled skutterudites towards composite thermoelectrics. *Npg Asia Materials* 2016;**8**(7):e285.
40. Tang Y, Chen S-w, Snyder GJ; Temperature dependent solubility of Yb in Yb–CoSb 3 skutterudite and its effect on preparation, optimization and lifetime of thermoelectrics. *J. Materiomics* 2015;**1**(1):75-84.
41. Williams CC, Wickramasinghe HK; Scanning Thermal Profiler. *Applied Physics Letters* 1986;**49**(23):1587-1589.
42. Zhao KY, Zeng HR, Xu KQ, et al.; Scanning thermoelectric microscopy of local thermoelectric behaviors in (Bi,Sb)(2)Te-3 films. *Physica B-Condensed Matter* 2015;**457**:156-159.

43. King WP, Kenny TW, Goodson KE, et al.; Atomic force microscope cantilevers for combined thermomechanical data writing and reading. *Applied Physics Letters* 2001;**78**(9):1300-1302.
44. Chui BW, Stowe TD, Kenny TW, et al.; Low‐stiffness silicon cantilevers for thermal writing and piezoresistive readback with the atomic force microscope. *Applied Physics Letters* 1996;**69**(18):2767-2769.
45. Vettiger P, Cross G, Despont M, et al.; The" millipede"-nanotechnology entering data storage. *Nanotechnology, IEEE Transactions on* 2002;**1**(1):39-55.
46. Eshghinejad A, Esfahani EN, Wang PQ, et al.; Scanning thermo-ionic microscopy for probing local electrochemistry at the nanoscale. *Journal of Applied Physics* 2016;**119**(20):205110.
47. Lefevre S, Volz S, Saulnier JB, et al.; Thermal conductivity calibration for hot wire based dc scanning thermal microscopy. *Review of Scientific Instruments* 2003;**74**(4):2418-2423.
48. King WP, Bhatia BS, Felts JR, et al.; Heated atomic force microscope cantilevers and their applications. *Annual Review of Heat Transfer* 2013;**16**(16).
49. Caillat T; Preparation and thermoelectric properties of Ir x Co 1− x Sb 2 alloys. *Journal of Physics and Chemistry of Solids* 1996;**57**(9):1351-1358.
50. Lal HB, Gaur K; Electrical-Conduction in Non-Metallic Rare-Earth Solids. *Journal of Materials Science* 1988;**23**(3):919-923.
51. Lee J, Wright TL, Abel MR, et al.; Thermal conduction from microcantilever heaters in partial vacuum. *Journal of Applied Physics* 2007;**101**(1):014906.

# Quantitative Nanoscale Mapping of Three-Phase Thermal Conductivities in Filled Skutterudites via Scanning Thermal Microscopy

**Supplementary Information**

**Sample microstructure**

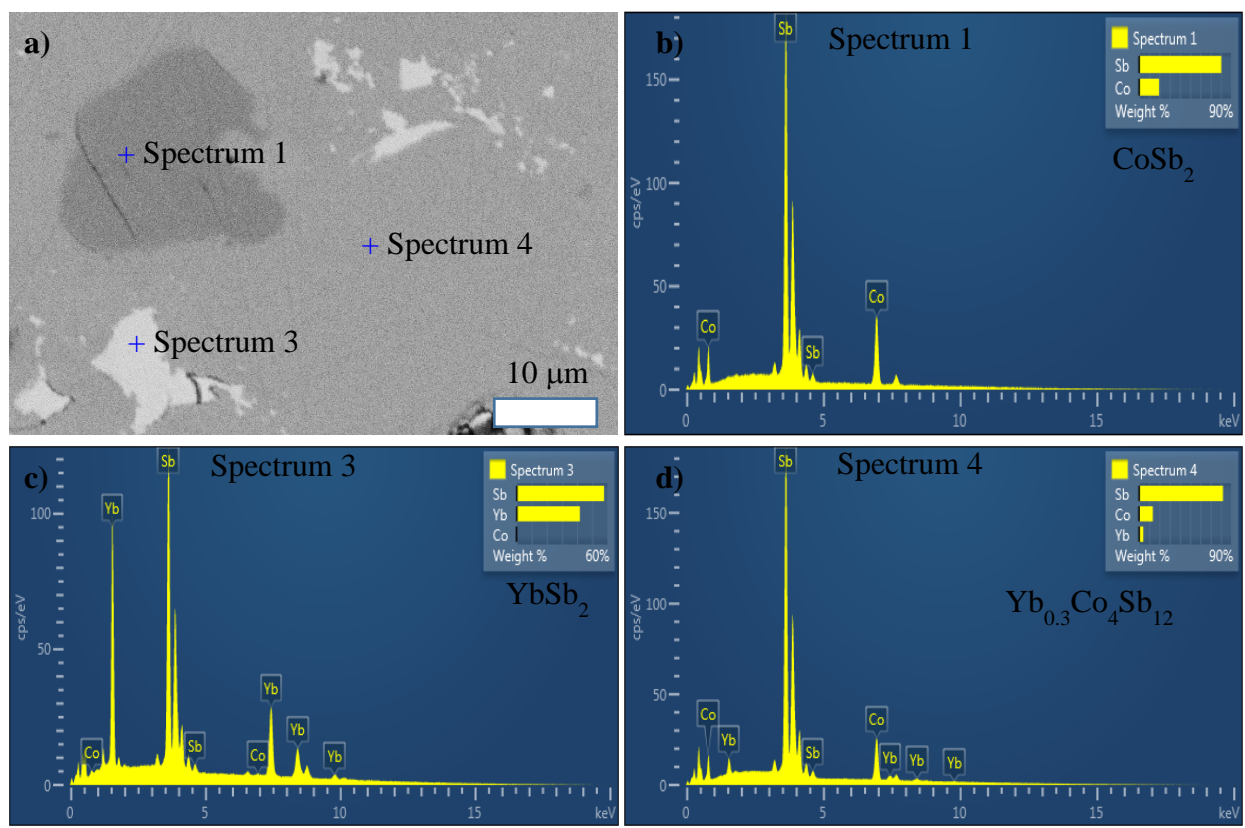

**Fig. S1** Energy dispersive X-ray spectrum EDS taken from different areas, using the Piont&ID mode in FEI Sirion XL30 SEM with high resolution Oxford EDS; (a) back scattering image; (b) EDS of the $CoSb_2$ phase; (c) EDS of the $YbSb_2$ phase; and (d) EDS of the $Yb_{0.3}Co_4Sb_{12}$ skutterudite phase.

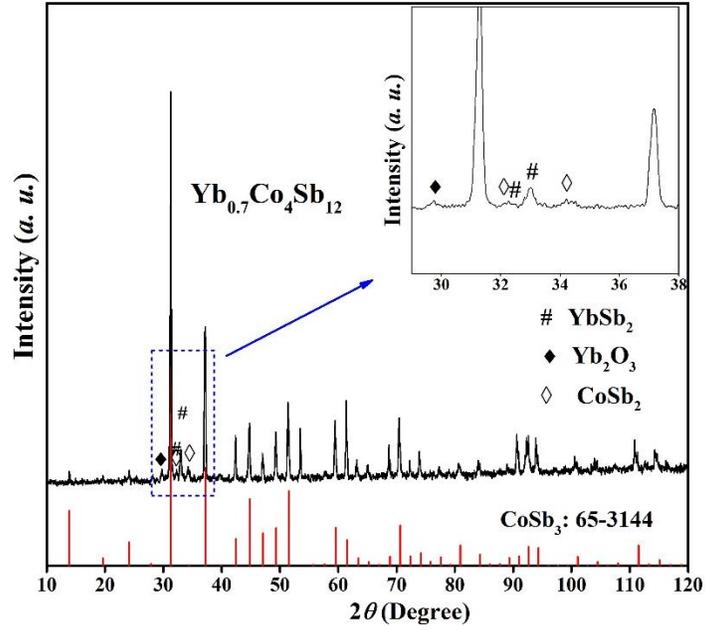

**Fig. S2** Powder X-ray diffraction pattern of $Yb_{0.7}Co_4Sb_{12}$; the inset clearly shows the existence of $YbSb_2$, $Yb_2O_3$, and $CoSb_2$ impurity phases.

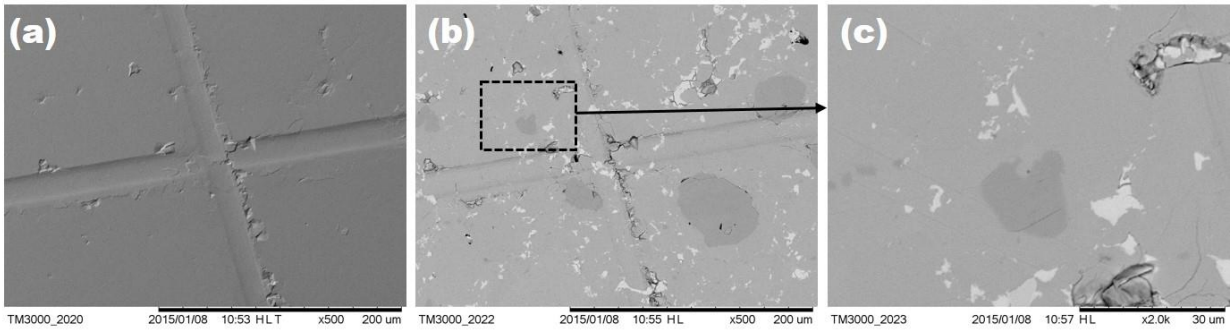

**Fig. S3** (a) SEM image of $Yb_{0.7}Co_4Sb_{12}$ sample with the artificial cross mark; (b) back scattered electron BSE image; and (c) zoom-in BSE image on the area of interest.

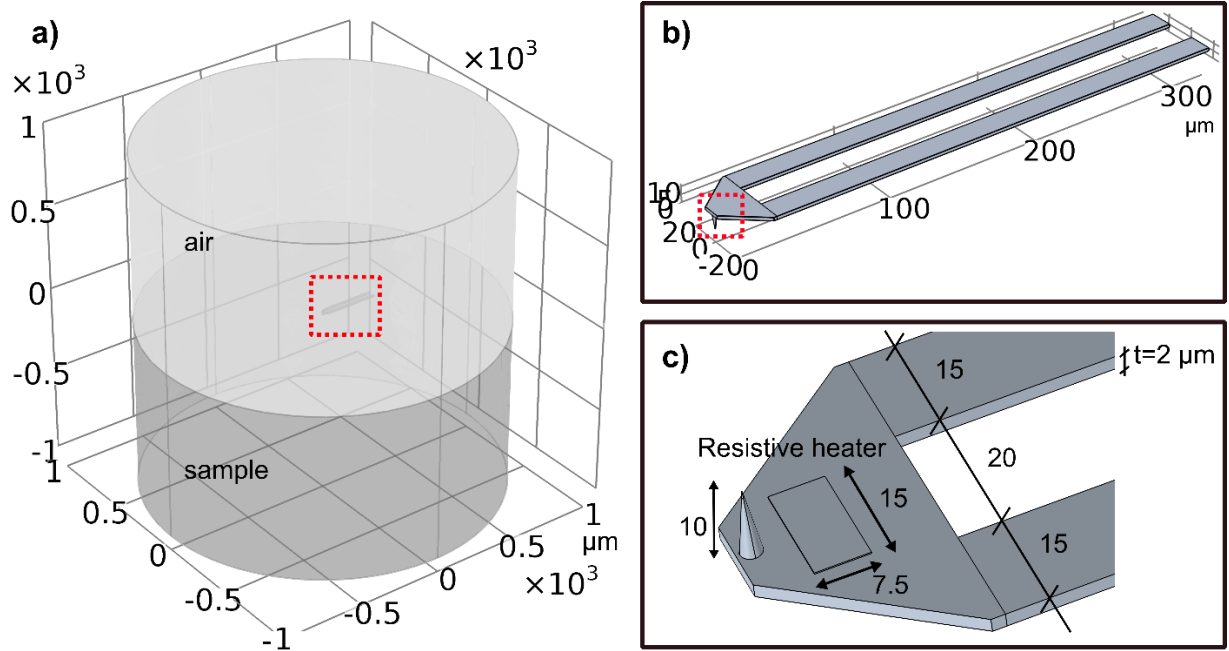

**Fig. S4** Finite element model of SThM; (a) sample, probe and the surround air; (b) zoom-in on the probe; and (c) further zoom-in on the probe apex and heater region; all dimensions shown are in micrometer and the probe tip radius is 20 nm. The dashed red boxes show the zoom-in area.

**Heat transfer simulations in COMSOL Multiphysics**

The heat conduction was defined on all domains (Fig. S4) using the heat transfer module based on the heat conduction equation:

$$\rho C_p \mathbf{u} \cdot \nabla T - \nabla \cdot (\kappa \nabla T) = Q,$$

where $\rho$, $C_p$, and $\kappa$ are the density, heat capacitance and thermal conductivity of each domain, respectively, $T$ is the absolute temperature, and $Q$ is the heat source. The heat source is only nonzero on the heater area (as shown in Fig.S4 (c)) through resistive Joule heating due to a constant voltage difference applied on the boundaries of the heater under Electric Currents module.

One important aspect of heat transfer in SThM is the thermal contact resistance between tip and sample. The contact resistance depends on the probe and sample geometry, their material properties, the nature of the contact and the contact area, and the contact force [1, 2]. In most of experimental measurements, the contact resistance is first determined by performing point-wise

SThM measurements on materials with known conductivity, and it is assumed that this contact resistance is not dependent on a small variation of sample thermal conductivity or changes in surface roughness [3, 4]. We used the same SThM probe throughout all the measurements, the contact force is kept strictly constant, and the geometry and the nature of the tip-sample contact remains the same. It should be also noted that we have intermediate variation of the thermal conductivity in our scanned sample, from 2 to 13 W/(m.K). Thus, the contact resistance can be treated as a constant [5].

We have implemented the thermal contact resistance $G_c$ in our model by defining heat fluxes at the upside and downside boundaries according to the relations [6]:

$$-\mathbf{n_d} \cdot (-\kappa_d \nabla T_d) = \frac{T_u - T_d}{G_c A_c} \quad \text{and} \quad -\mathbf{n_u} \cdot (-\kappa_u \nabla T_u) = \frac{T_d - T_u}{G_c A_c},$$

where **n** refers to interfacial normal and the parameters at upside and downside are denoted by subscript. We used a thermal resistance $G_c$ of $1.0 \times 10^8$ K/W for the contact $A_c$ with radius 20 nm, which was measured using similar type of SThM probe in a previous study [7].

We first examine the effect of the thermal contact resistance by comparing the temperature and heat flux distributions along the tip-sample interface. As it is shown in Fig. S5, without considering the losses, the temperature should be continuous at the interface, while by defining the contact resistance, the temperature and heat flux drops significantly at the tip-sample junction. The question here is whether this temperature drop affects the temperature of the probe heater and its electrical resistance or not. For this purpose, we compare the resistive heater temperature and its electrical resistance as a function of the sample thermal conductivity, with and without considering the contact resistance. It is observed that although the tip temperature drops more than 20 K, the temperature and the electrical resistance of the resistive heater are not significantly affected, as shown in Figs. S5 and S6. The change in the electric resistance after considering the thermal resistance for a sample thermal conductivity of 12 W/(m.K) is less than 0.1%.

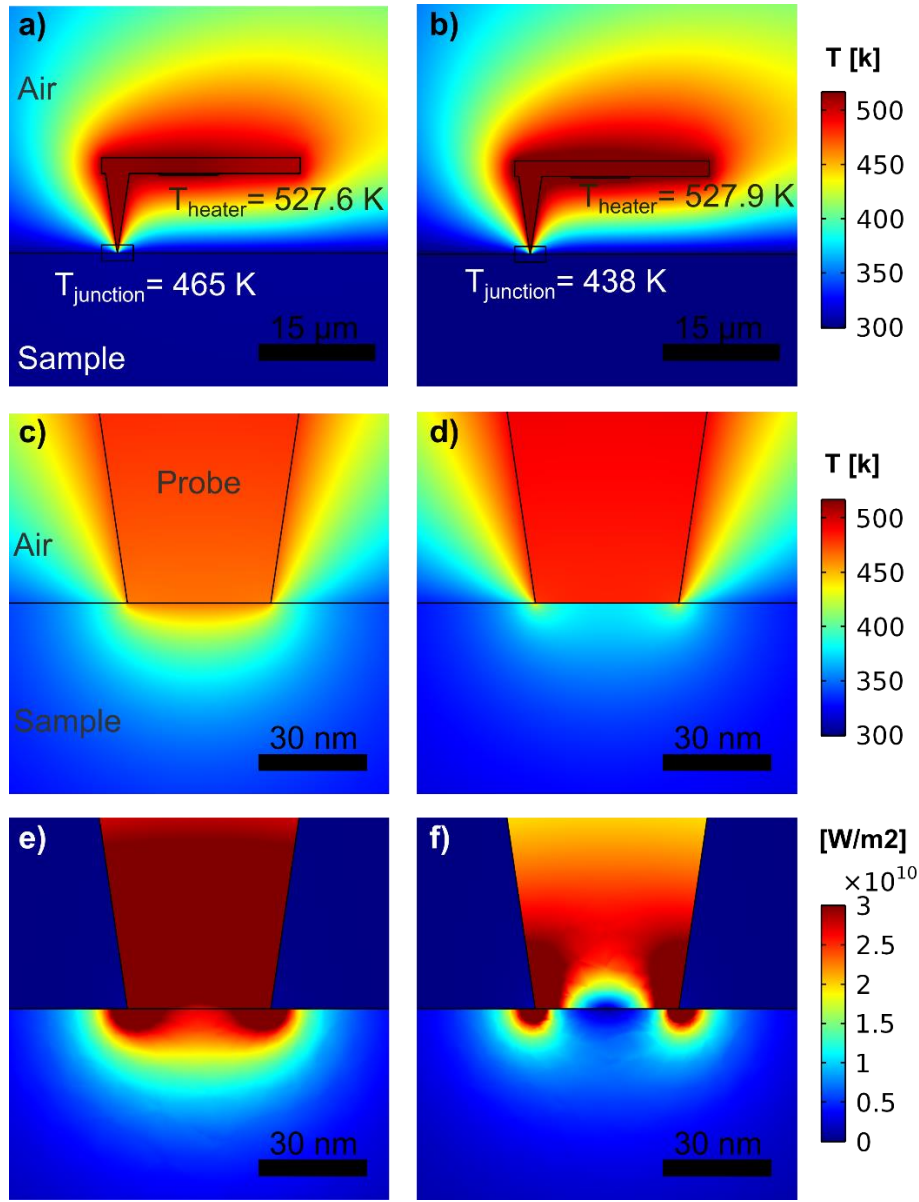

**Fig. S5** Numerical simulation of SThM; the cross section temperature distributions along the probe (a,b) and near the probe-sample junction (c,d), and conductive heat flux distribution (e,f) near the probe-sample junction with. Thermal contact resistance was ignored on the left side (a,c,e) but considered in the right wide (b,d,f). In the simulation, the thermal conductivity of the sample is taken to be 5 W/(m.K), and the contact resistance is taken to be $1.0 \times 10^8$ K/W.

It is also observed in Fig. S6 that SThM measurement is sensitive to sample thermal conductivity change, yet insensitive to contact resistance change. From the top row of the figure, it is observed that with a contact resistance of $1.0 \times 10^8$ K/W, the maximum reduction in heater temperature and resistance is only 0.11% and 0.03%, respectively, occurring at thermal conductivity of 12 W/(m.K), while thermal conductivity changing from 2 W/(m.K) to 12 W/(m.K) resulting in a reduction of heater temperature and resistance around 1.15% and 0.36%, which is much more substantial. This is more evident in the bottom row, showing that the change in heater temperature and resistance is insignificant when the contact resistance varies by four orders of magnitude.

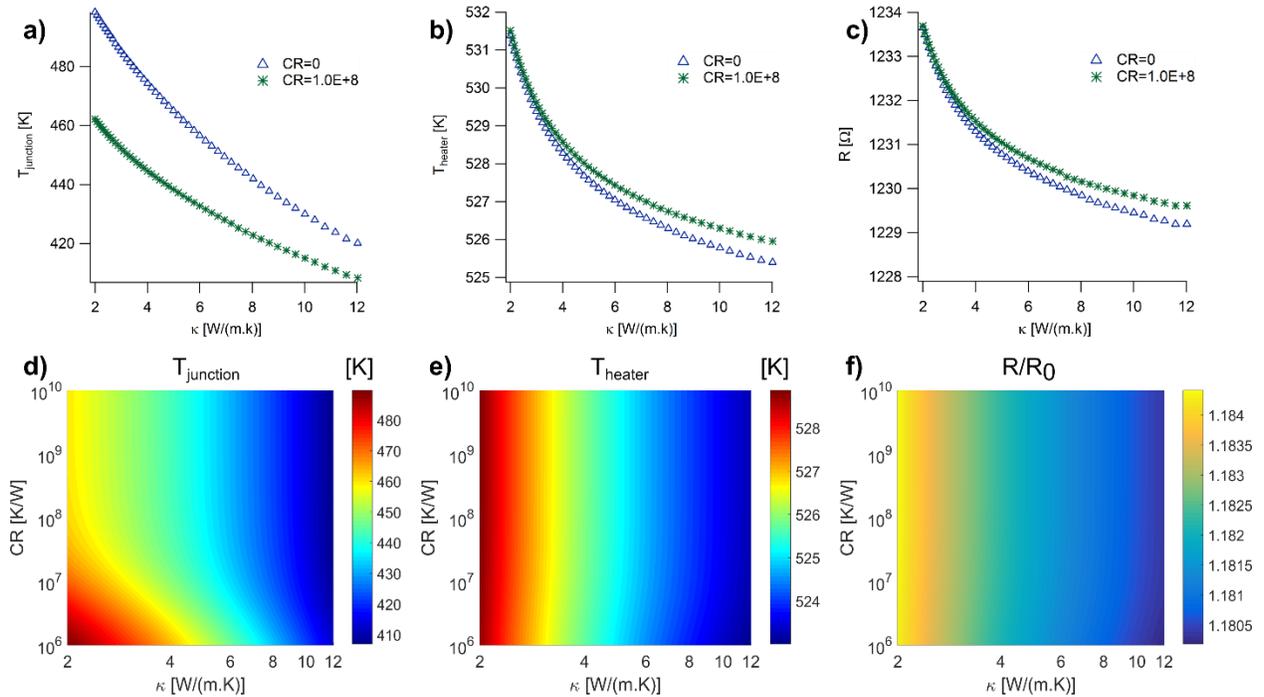

**Fig. S6** Sensitivity of SThM measurements as a function of sample thermal conductivity $\kappa$ and contact resistance CR as revealed by numerical simulations; (a,d) temperature at probe-sample junction; (b,e) average temperature of the heater; and (c,f) the electrical resistance corresponding to the heater temperature. The top row show the variation with respect to thermal conductivity, while the bottom row show the two-dimensional mapping with respect to thermal conductivity and contact resistance.

Another issue is the effect of convection and radiation, which can be ignored as shown by the following analysis. In the simulation, the natural convection in the air is considered and solved by using a Laminar Flow module in COMSOL by solving the full Navier-Stokes equations considering the pressure shift:

$$\rho(\mathbf{u} \cdot \nabla \mathbf{u}) = -\nabla p + \nabla \cdot \left(\mu(\nabla \mathbf{u} + (\nabla \mathbf{u})^T) - \frac{2}{3}\mu(\nabla \cdot \mathbf{u})\mathbf{I}\right) + (\rho - \rho_0)g,$$

where $\mathbf{u}$ and $p$ are the fluid velocity and pressure fields, $\rho_0$ is the reference density of the air in room temperature, $\mu$ is the dynamic viscosity, $\mathbf{I}$ is the identity matrix, and $g$ is the acceleration due to gravity. The density and dynamic viscosity are temperature and pressure dependent which couples the Laminar Flow and Heat Transfer modules. The boundaries between air and solids are defined as no-slip walls and the external boundaries are defined as open boundaries, *i.e.*, the normal stress on boundary is zero.

The surface-to-surface radiation is defined on the bottom part of the probe and the top surface of the sample by defining a radiation flux between surface $i$ to $j$, as:

$$Q_{ij} = \sigma A_i \epsilon_i (T_i^4 - T_j^4),$$

where $\epsilon_i$ is the surface emissivity of surface $i$ conservatively assumed to be 0.6 for all surface, $A_i$ is the surface area, and the Stefan-Boltzmann constant $\sigma = 5.67 \times 10^{-8}$ [W. m$^{-2}$. K$^{-4}$].

The radiative and conductive heat fluxes on the upper surface of the sample are compared in Fig. S7, where we observed a conductive heat flux orders of magnitude higher than the radiative one. The total radiative heat transfer on the sample surface was less than 5 µW while the conductive one was in order of 9 mW. Similar simulation without considering the effect of radiation resulted in probe temperature difference of less than 0.01 K, reassuring that the radiation can be completely ignored.

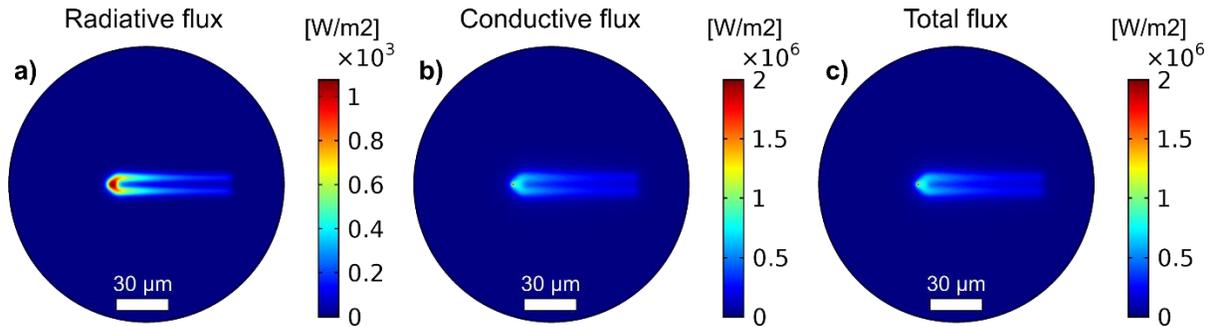

**Fig. S7** Simulations of SThM with radiative and conductive heat transfer. Comparison of the (a) radiative, (b) conductive, and (c) total heat flux magnitude mappings on the sample surface (zoom-in). The thermal conductivity of the sample was 5 W/(m.K).

To investigate the effect of natural convection, we compare the conductive and convective heat transfer mechanisms where the thermal probe is operated in contact with a sample of thermal conductivity 5 W/(m.K). The results are illustrated in Fig. S8, where the fluid velocity was observed to be less than 0.003 m/sec, the convective heat flux in the air is orders of magnitude smaller than of the conductive one, and the total heat transfer via natural convection was ~40 µW while the total conductive heat transfer was in order of 9 mW. Furthermore, we performed identical simulations without natural convection, where the difference in the probe temperature was less than 0.75 K and the variation in the electrical resistance of the probe was less than 0.1%. Therefore, it is also safe to assume that the convection does not influence the SThM results.

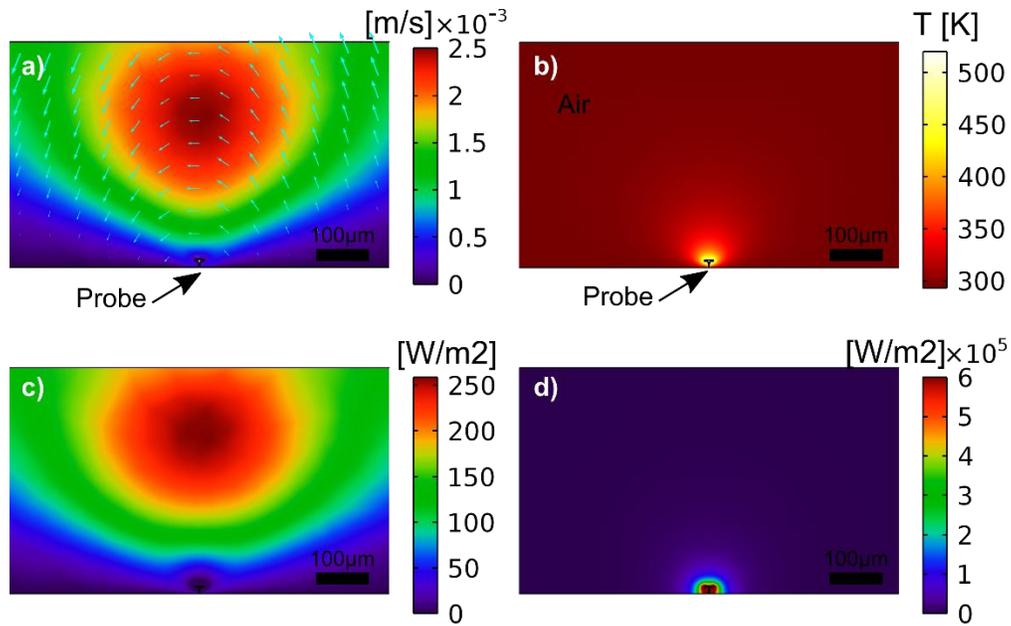

**Fig. S8** Simulation of SThM considering natural convection and heat conduction; (a) fluid velocity magnitude and field; (b) temperature distribution; (c) convective and (d) conductive heat flux distributions in the air-gap while the probe is contact with the sample ($\kappa = 5$ W/(m.K)).

**Macroscopic thermoelectric properties**

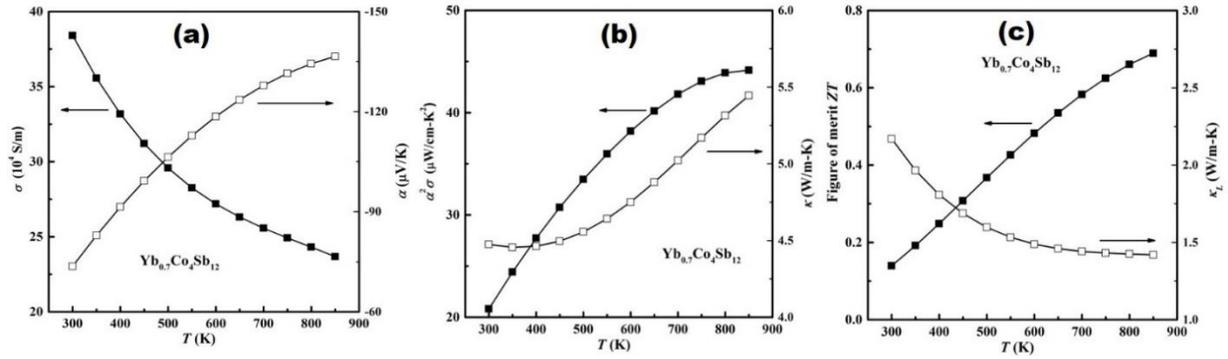

**Fig. S9** Thermoelectric properties of $Yb_{0.7}Co_4Sb_{12}$ sample; (a) electrical conductivity $\sigma$ and Seebeck coefficient $\alpha$; (b) power factor $\alpha^2\sigma$ and thermal conductivity $\kappa$; and (c) $ZT$ and lattice thermal conductivity $\kappa_L$.

**Transient behavior of heat transfer**

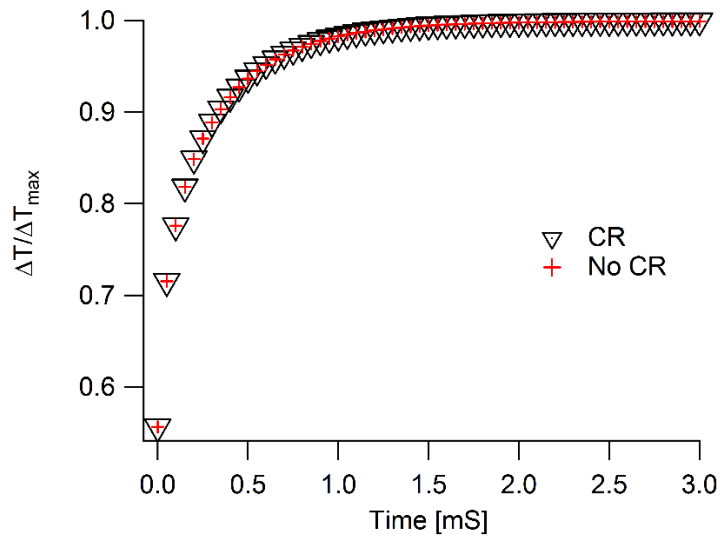

**Fig. S10** Time-dependent analysis of thermal probe in touch with a sample having $\kappa = 5\,\mathrm{W/m\cdot K}$ with and without considering the thermal contact resistance; The transient time for the thermal probe to reach steady-state condition is around 1.2mSec.

**Materials parameters**

Table S1 Material properties used in FE simulations

| Material | Thermal conductivity $k$ [$\frac{W}{mK}$] | Heat Capacity $C_p$ [$\frac{J}{kg*K}$] | Density [$\frac{kg}{m^3}$] |
|---|---|---|---|
| Silicon (Cantilever) | 130 | 700 | 2329 |
| Sample | 3.2-15 | 300 | 7000 |
| Air | 0.025645 | 1010.19 | 1.2 |

**Table S2** Thermal conductivity of calibration samples

| Material | Thermal conductivity $k\ [\frac{W}{mK}]$ |
|---|---|
| BiSbSe$_3$ | 0.66 |
| Te | 1.41 |
| Yb$_{0.25}$Co$_4$Sb$_{12}$ | 3.17 |
| Bi | 7.45 |
| CoSb3 | 10 |
| FeNbSb | 16.6 |
| YbSb2 | 16.9 |
| Sb | 22.3 |
| Pb | 34.3 |
| Sn | 71.3 |
| Co | 80.8 |

**Table S3** Transport properties of four relevant phases at 300 K

| Phases | $\sigma$ (S/m) | $\kappa$ (W/m-K) | $\alpha$ (μV/K) | Remark |
|---|---|---|---|---|
| $Yb_{0.3}Co_4Sb_{12}$[†] | $2.3\times10^5$ | 3.2 | -138 | Polycrystal |
| $CoSb_2$[#] | $3.0\times10^5$ | 11.8 | 26 | Single-crystal |
| $YbSb_2$ | $7.9\times10^6$ | 15.0 | 15 | Polycrystal |
| $Yb_2O_3$ | $2.4\times10^2$ [*] | - | - | no $\kappa$ value found |

[†] [8], [#] [9], [*] [10]

**References**

1. Rojo MM, Martin J, Grauby S, et al.; Decrease in thermal conductivity in polymeric P3HT nanowires by size-reduction induced by crystal orientation: new approaches towards thermal transport engineering of organic materials. *Nanoscale* 2014;**6**(14):7858-65.

2. Shi L, Majumdar A; Thermal transport mechanisms at nanoscale point contacts. *Journal of Heat Transfer-Transactions of the Asme* 2002;**124**(2):329-337.

3. Hinz M, Marti O, Gotsmann B, et al.; High resolution vacuum scanning thermal microscopy of Hf O 2 and Si O 2. *Applied Physics Letters* 2008;**92**(4):043122.

4. Zhang YL, Hapenciuc CL, Castillo EE, et al.; A microprobe technique for simultaneously measuring thermal conductivity and Seebeck coefficient of thin films. *Applied Physics Letters* 2010;**96**(6):062107.

5. Pumarol ME, Rosamond MC, Tovee P, et al.; Direct nanoscale imaging of ballistic and diffusive thermal transport in graphene nanostructures. *Nano Lett* 2012;**12**(6):2906-11.

6. Swartz ET, Pohl RO; Thermal-Boundary Resistance. *Reviews of Modern Physics* 1989;**61**(3):605-668.

7. King WP, Bhatia BS, Felts JR, et al.; Heated atomic force microscope cantilevers and their applications. *Annual Review of Heat Transfer* 2013;**16**(16).

8. Wang SY, Salvador JR, Yang J, et al.; High-performance n-type YbxCo4Sb12: from partially filled skutterudites towards composite thermoelectrics. *Npg Asia Materials* 2016;**8**(7):e285.

9. Caillat T; Preparation and thermoelectric properties of Ir x Co 1− x Sb 2 alloys. *Journal of Physics and Chemistry of Solids* 1996;**57**(9):1351-1358.

10. Lal HB, Gaur K; Electrical-Conduction in Non-Metallic Rare-Earth Solids. *Journal of Materials Science* 1988;**23**(3):919-923.